# Mid-Infrared Spectroscopy and Microscopy of Subcellular Structures in Eukaryotic Cells with Atomic Force Microscopy - Infrared Spectroscopy.

*Luca Quaroni,\* Katarzyna Pogoda, Joanna-Wiltowska Zuber, Wojciech Kwiatek*

*Department of Experimental Physics of Complex Systems, Institute of Nuclear Physics, Polish Academy of Sciences, PL-31342, Kraków, Poland*

**Abstract**

Atomic Force Microscopy – Infrared (AFM-IR) spectroscopy allows spectroscopic studies in the mid-infrared spectral region with a spatial resolution better than 50 nm. We show that the high spatial resolution can be used to perform spectroscopic and imaging studies at the subcellular level in fixed eukaryotic cells. We collect AFM-IR images of subcellular structures that include lipid droplets, vesicles and cytoskeletal filaments, by relying on the intrinsic contrast from IR light absorption. We also obtain AFM-IR absorption spectra of individual structures. Most spectra show features that are recognizable in the IR absorption spectra of cells and tissue obtained with FTIR technology, including absorption bands characteristic of phospholipids and polypeptides. We also observe sharp spectral features that we attribute to the nonlinear photothermal response of the system and we propose that they can be used to perform sensitive spectroscopy on the nanoscale.

## Introduction

Infrared absorption spectroscopy in the mid-infrared (mid-IR) spectral region, about 400 cm$^{-1}$ - 4000 cm$^{-1}$ (25 µm - 2.5 µm), is frequently used for the characterization of biological matter at the molecular level. Molecular species strongly absorb mid-IR radiation, meaning that the technique is generally applicable to a large variety of samples, ranging from purified biomolecules to cells and tissue,[1] including living cells.[2] Mid-IR light absorption provides information on composition and structural properties of the samples, making it valuable in the study of structure-function relationships, reaction mechanisms and dynamics, [3] and in medical diagnostic applications. [4] The technique can be implemented in a microscopy configuration [5] and allows spectromicroscopy and imaging experiments to be performed on samples as small as individual eukaryotic cells. [6] [7]

A major limitation of mid-IR microscopy is imposed by the use of far-field optics, which constrain the spatial resolution to the limits set by diffraction, of the order of the wavelength of light. This corresponds to a few micrometers in the mid-IR region, comparable to the size of most eukaryotic cells. [8] Because of these constraints, mid-IR imaging can provide only limited information about subcellular structure. In most eukaryotic cells the largest structures that can be resolved are the nucleus and larger vacuoles. In smaller eukaryotic cells, such as some neuronal cells, subcellular imaging is not viable. Spectromicroscopy mid-IR experiments on eukaryotic cells are mostly performed at the single cell level, averaging the information over the whole cell. Most organelles, such as mitochondria, chloroplasts or lysosomes, or the membranes of the endoplasmic reticulum and the Golgi apparatus, are too small to be probed selectively. Despite the novel, rich and sometimes unique information potentially available on cellular biochemistry, limits in spatial resolution have been a major obstacle to the implementation of cellular studies by mid-IR absorption spectroscopy.

In recent years, efforts have been under way to circumvent the resolution constraints set by diffraction by performing mid-IR absorption spectroscopy with near-field optics. Several optical configurations have been demonstrated, [9] including transmission-mode and scattering mode Scanning Near-field Optical Microscopy (SNOM and s-SNOM respectively, also SNIM for the latter), [10] [11] and Photo-Thermal Infrared detected by a temperature probe [12] or cantilever deflection (PTIR or AFM-IR). [13, 14] Spatial resolution lower than the wavelength of the probing light, and in some cases lower than 20 nm, [15, 16] has been reported for this optical configuration.

An AFM-IR experiment uses the deflection of an AFM tip to detect local photothermal sample expansion following absorption of light. [17] The expansion excites normal modes of the AFM cantilever, the oscillation of which is measured by the movement of the AFM laser on the four-quadrant detector used to monitor AFM experiments. The amplitude of the oscillation is a measure of the amount of light absorbed by the sample. In experiments with pulsed light sources, the amplitude of the ringdown of the oscillation or of one of the peaks in its resonance spectrum is used to record indirectly the absorption spectrum of the sample. Therefore, scanning the wavelength of the source laser provides an absorption spectrum of the sample in the contact location of the AFM tip. In alternative, scanning the sample surface as in a standard AFM experiment, while keeping the exciting wavelength fixed, provides a map of the absorption at that wavelength throughout the sample.

The performance of AFM-IR is characterized by three concomitant features. First, the accessible spatial resolution is comparable to the size of the AFM tip. Second, absorption spectra obtained by AFM-IR measurements are a function of the imaginary part of the refractive index of the sample, with no contribution from the real part. Therefore, an AFM-IR spectrum of the sample generally reproduces the transmission spectrum of the same material measured with a conventional Fourier Transform Infrared (FTIR) instrument. This contrasts with measurements performed with other near-field configurations, where both the real and imaginary part of the refractive index of the sample can contribute to spectral bandshape. The capability of AFM-IR to obtain the same spectra provided by a bulk measurement makes this a technique of choice for performing spectroscopy on small samples. Thirdly, an AFM-IR experiment consists of parallel and concomitant AFM and spectroscopic measurements performed on the same sample, thus providing spectroscopic, topographic and mechanical information in one single run. [13]

Since its inception, AFM-IR has been applied to the study of soft matter, including biological samples, such as protein aggregates and single cells. The achievable resolution has been particularly useful in the study of single prokaryotic cells, the size of which is of the order of the micrometer and inaccessible to far-field mid-IR measurements. [18-20] Despite the potential of the technique for the study of cellular biochemistry, relatively few studies have been performed on eukaryotic cells. Several of these investigations were aimed at detecting the subcellular distribution of exogenous species, such as carbonyl-metal complexes, [21] drugs [22] or nanoparticles.[23] Some investigations addressed the study of intrinsic biomolecules in eukaryotic cells, including triglyceride vesicles in algal and yeast cells, [20] polysaccharide distribution in live *Candida* hyphae, [24] nuclear localization in cancer cells [25] and protein distribution in HeLa cells. [26]

In this work, we apply AFM-IR to the study of fixed fibroblast cells. We demonstrate the possibility to study subcellular structures such as cytoskeletal components, vesicles and micelles using both spectromicroscopy and imaging experiments.

## Experimental

AFM-IR measurements were performed on a nanoIR2 instrument (Anasys, Santa Barbara, CA, USA) working in contact mode using PR-EX-nIR probes with nominal tip diameter of 30 nm and eigenfrequency equal to 12.79 ±0.64 kHz. The photothermal induced deflection of the cantilever was measured by the movement of a laser beam reflected from its surface and recorded by a four-quadrant detector as a sinusoidal decay signal. The resonance spectrum of the cantilever was extracted from the oscillatory decay using the Fast Fourier-Transform (FFT) algorithm.

An OPO (Optical Parametric Oscillator) laser was used as the excitation source. For the measurement of extended spectra, the laser was scanned from 850 cm$^{-1}$ to 3600 cm$^{-1}$ in 2 cm$^{-1}$ steps. The plane of polarization was at 0 degrees. Power was set at 10% of the maximum (about 8 mW) and 1024 measurements were co-averaged for each spectral point. The maximum peak-to-peak amplitude of the oscillatory decay was used for recording AFM-IR spectra.

For the measurement of maps, the laser was set at the selected wavenumber while the AFM tip was scanned over the sample in contact mode. The contact mode scan was performed using a scan rate of 0.1 Hz, a 256-pixel resolution in the X and Y direction, and turning on the feedback loop of the Z scanner. The plane of polarization was at 0 degrees. Power was set at 19% of the maximum and 8 or 32 measurements were co-averaged for each spectral point. The maximum peak-to-peak amplitude of the oscillatory decay was used for recording AFM-IR maps.

Detroit 551 human skin fibroblasts were obtained from the European Collection of Authenticated Cell Cultures (ECACC, Porton Down, UK) and measured onto CaF2 optical substrates after fixation.

Cells were treated with fluorescent stains to be used for fluorescence microscopy. Bodipy 493/503, 1:200, (Molecular Probes, Eugene, OR, USA) was used to label neutral lipids and lipid droplets, rhodamine-phalloidin, 1:200, (Molecular Probes, Eugene, OR, USA) to label actin cytoskeleton filaments and Hoechst, 1:1000, (Molecular Probes, Eugene, OR, USA) to label cell nuclei. Cells were visualized using an AxioImager Z2 fluorescence microscope (Carl Zeiss, Oberkochen, Germany) equipped with a Plan-Apochromat (10x/0,45NA) objective and a Plan-Apochromat (63x/1,40NA) oil immersion objective. Fluorescence images were analyzed using Image J. [27]

## Results

We have used Detroit 551 human skin fibroblasts as a model system for our studies. Fibroblast cells are an ideal sample because they grow as an adherent phenotype on a variety of substrates, including optical substrates that are transparent to mid-IR radiation, such as $CaF_2$. Adherent cells can be measured repeatedly with reduced problems due to sample displacement by the AFM tip. Finally, coverage can be controlled to provide isolated cells, cell clusters or a cell monolayer. In this study, we performed AFM-IR measurements on a monolayer of adherent cells grown to full confluency.

Fluorescent images of fixed Detroit 551 fibroblasts are shown on Figure 1. The images show the presence of lipid droplets and actin filaments in paraformaldehyde fixed cells.

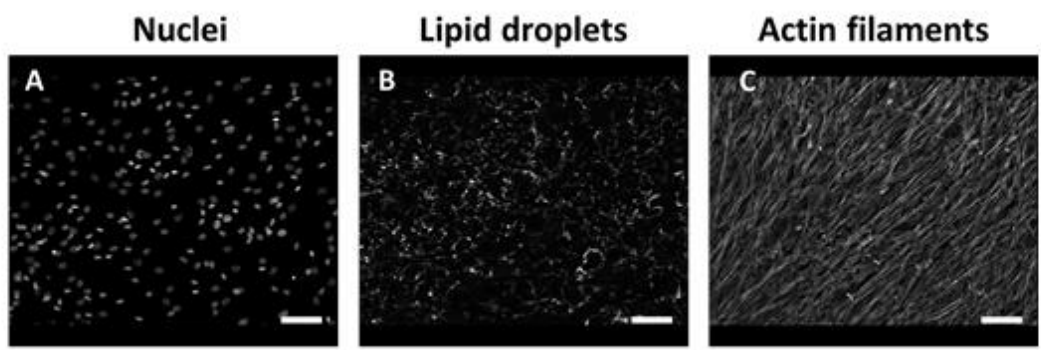

*Figure 1.* *Comparison of fluorescent images obtained by staining nuclei (A), lipid droplets (B) and actin filaments (C) of fixed Detroit 551 fibroblasts.*

Single point spectral measurements and single wavelength images shown in this work were all collected on the same sample, together with AFM images. The CaF$_2$ substrate used for cell growth is transparent to light in the Mid-IR spectral region above 900 cm$^{-1}$, and allows using the full spectral range accessible to the OPO laser. The area shown in Figure 2 corresponds to part of three cells in contact, lying side by side along their major axis. Part of a nucleus can be seen on the left edge of the field of view. Supplementary Figure S1 shows a wider scan of the surrounding portion of the sample, allowing an extended view of the sample and clear identification of cell layout and nuclear positions.

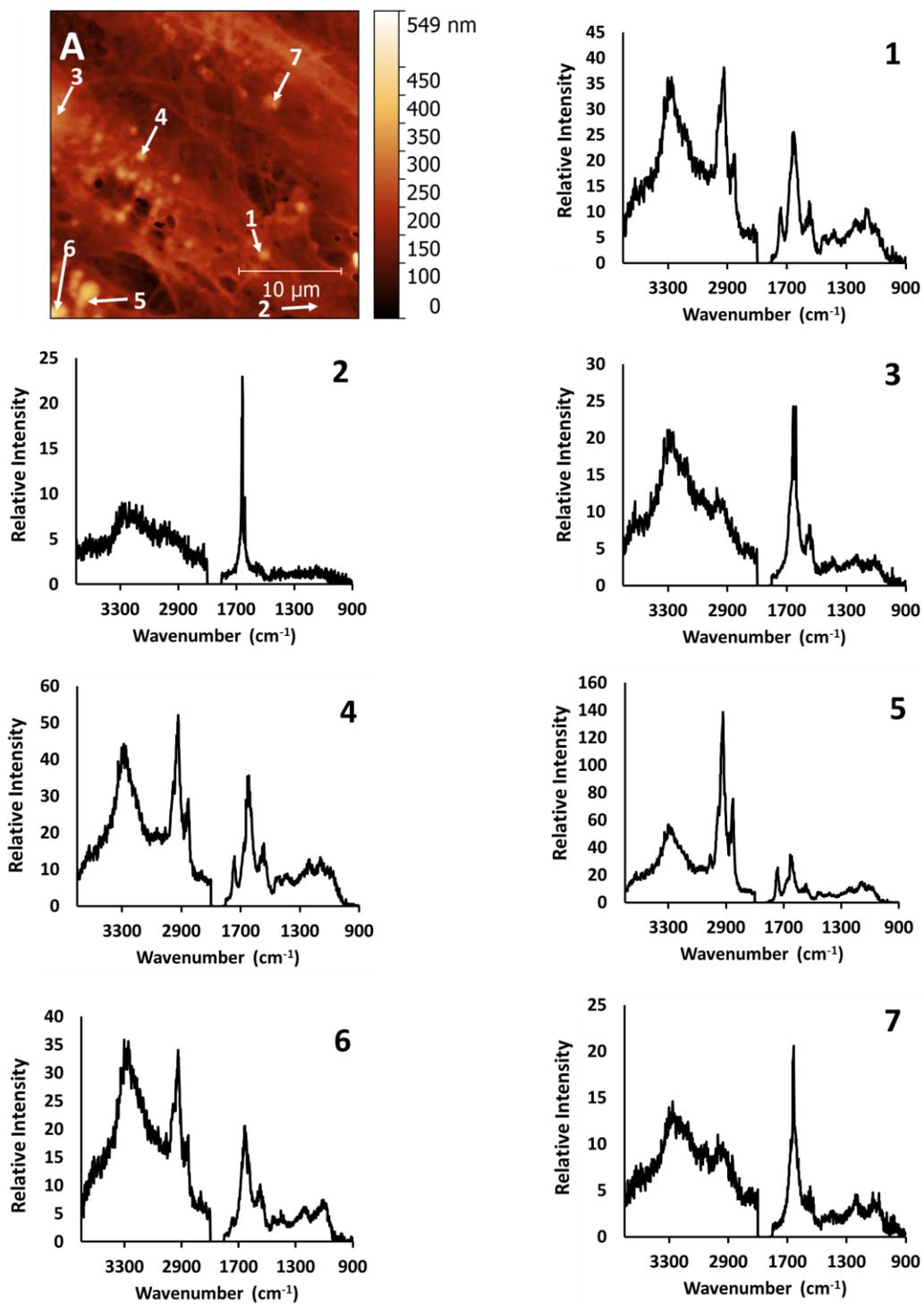

*Figure 2.* AFM-IR spectra measured at specific locations of fixed fibroblasts. A – Height AFM image. 1 - 7 – Single point spectra measured at the marked locations in panel A.

The AFM height plot in Figure 2 (panel A) provides a topographic description of the sample. When the cell is fixed, it becomes a hard object and can be routinely imaged using AFM contact mode. Such measurement reveals the features exposed at the cell surface, mostly highly organized fibrous structures, between 100 nm and 1000 nm, that create the cell actin cytoskeleton. Cell membranes are usually more difficult to visualize and the topography image is mostly dominated by stiff protein fibers. Paraformaldehyde fixation results mainly in crosslinking of amine groups of proteins but it is also recognized as an effective method for preserving lipid droplet structure. Lipid droplets, usually occupying the cell interior, can be seen in the AFM topography image.

AFM-IR spectra of the mid-IR spectral range were recorded at selected positions of the cells and are shown Figure 2. The corresponding measurement positions are marked on the Height image of Figure 2 (panel A). Panels 1 to 7 display the spectra recorded at the numbered locations. The spectral range was limited to the 950 cm$^{-1}$ - 1800 cm$^{-1}$ and 2700 cm$^{-1}$ - 3600 cm$^{-1}$ intervals, to exclude spectral regions in which no absorption from the sample occurs.

The infrared absorption spectra of eukaryotic cells have been extensively described using FTIR microscopy with far-field optics. FTIR absorption spectra of single fixed eukaryotic cells are qualitatively similar for most cells and are dominated by a contribution from the main classes of biological macromolecules, mostly proteins, polysaccharides and nucleic acids, and phospholipids. [6]

We use the AFM Height map (Figure 2A) to orient ourselves on the sample and select specific locations for the measurement of single AFM-IR spectra. The positions selected for measurement are marked with numbers 1-7. The resulting spectra are shown in Figure 2, panels 1-7. The selected subcellular locations display spectra with bands that overall are similar to the ones observed in far-field FTIR spectra of whole cells. A common spectral pattern is given by bands that are characteristic for long chain acyl lipids, such as the strong doublet at 2920 cm$^{-1}$ and 2855 cm$^{-1}$, arising from the stretching modes of C-H bonds in CH$_2$ groups of the acyl chains. Additional weaker bands that arise from lipid vibrations are also observed. A weak but sharp band can be seen at 3010 cm$^{-1}$, corresponding to the stretching mode of C-H bonds on unsaturated C=C bonds. The weak shoulders at 2958 cm$^{-1}$ and 2895 cm$^{-1}$ are assigned to the stretching C-H modes of the terminal CH$_3$ groups in the acyl chains. The band at 1740 cm$^{-1}$ is due to the stretching of the C=O bond of the ester carbonyls in the phospholipid headgroups. Additional bands can be seen around 1080 cm$^{-1}$ - 1090 cm$^{-1}$ and 1220 cm$^{-1}$ - 1240 cm$^{-1}$, which are characteristic of the phosphate headgroups of phospholipids, and around 1380 cm$^{-1}$ - 1390 cm$^{-1}$ and 1430 cm$^{-1}$ - 1450 cm$^{-1}$, which are associated to bending modes of the CH$_2$ groups. The relative intensity of these bands is qualitatively similar to the one observed in FTIR measurements of bulk lipids. The only difference is given by the intensity of the peak at 2920 cm$^{-1}$, which appears to be relatively stronger than the one at 2855 cm$^{-1}$ compared to bulk isotropic samples.

The spectra of Figure 2 show that the band pattern characteristic of acyl lipids is associated to several particle-like structures observed in the sample (Positions 1, 4-6). It is weak or negligible when measured in other locations (Positions 2, 3, 7).

Some of the strongest spectral contributions in the FTIR spectra of cells are provided by the Amide I and Amide II bands, around 1650 cm$^{-1}$ and 1550 cm$^{-1}$ respectively. The bands arise from the amide groups of polypeptides. In purified proteins, these bands are used to quantify the relative abundance of specific secondary structure components in proteins. [3] Additional bands from proteins are present in the 1200 cm$^{-1}$ - 1400 cm$^{-1}$ spectral region (Amide III band, weak) and in the 3100 cm$^{-1}$ - 3300 cm$^{-1}$ region (Amide A and Amide B bands, strong).

A set of bands attributable to proteins can be observed in all the AFM-IR spectra recorded in the sample, including several of the particles (Figure 2, Positions 1, 4-7) and the nucleus (Position 3). Bands can be clearly seen in positions characteristic of Amide I, Amide II and Amide A components. The weaker Amide III overlaps with bands from other cellular components, such as lipids and cannot be resolved from them. The relative intensity of the bands is comparable to the one observed in bulk FTIR samples, with the Amide I being the stronger band. In comparison with FTIR spectra of bulk isotropic samples, in AFM-IR spectra the ratio of intensity between the Amide I and Amide II samples appears to be farther skewed towards the Amide I.

The spectrum in Position 3 (Figure 2) has been recorded in the nuclear position. It is dominated by protein Amide I and Amide II bands, with little contribution from the absorption of phospholipid bands. These features have been previously reported in the FTIR spectra of the nuclear region of eukaryotic cells and presumably arise from the abundance of chromatin and the lack of stacked membrane layers within the nucleus itself. [28]

The spectrum of the particle in Position 7 is dominated by contributions from Amide bands. However, despite the positive profile in the AFM Height map, the intensity in this position in the 1650 cm$^{-1}$ map is lower when compared to the surrounding region, suggesting that it is an empty cavity surrounded by a protein shell, presumably held together by cross-linking due to the fixation process.

Oligo and polysaccharides and the ribose chains of nucleic acids give rise to complex absorption multiplets in the 1000 cm$^{-1}$ - 1100 cm$^{-1}$ range, mostly due to vibrations of C-O bonds. In the spectra of Figure 2 varying contributions are observed in this spectral region, overlapping with those from phosphate and phosphate esters. Due to the complexity of the 1000 cm$^{-1}$ - 1100 cm$^{-1}$ region, it is not possible to visually identify band patterns specific to polysaccharides, although they are expected to provide a contribution to overall intensity.

In addition to absorption from ribose, nucleic acids also contribute bands from the absorption of the phosphate backbone around 1080 cm$^{-1}$ and 1250 cm$^{-1}$, and from the carbonyl and imide groups of the nucleic bases, around 1660 cm$^{-1}$ - 1680 cm$^{-1}$. As for polysaccharides, we cannot isolate specific contributions from nucleic acids by visual inspection of the spectrum because of the overlap with bands from other abundant components.

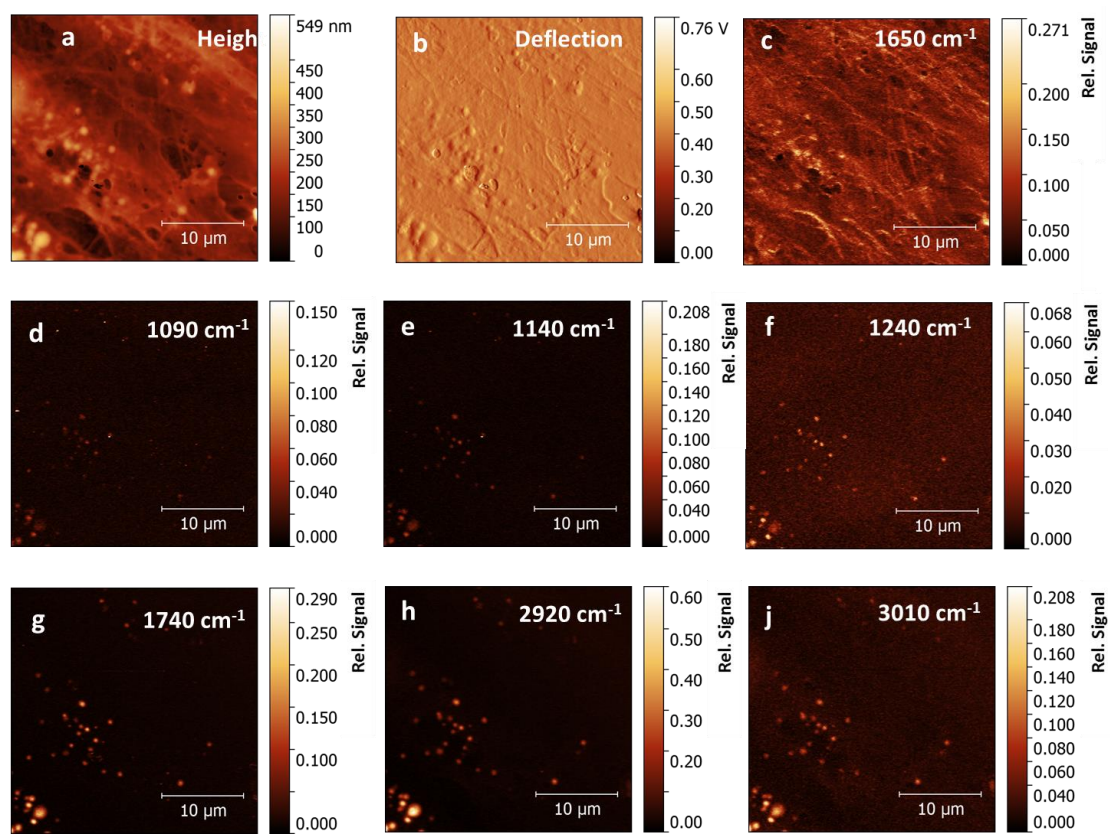

*Figure 3.* AFM-IR measurements on fixed fibroblasts. a – Height AFM image. b – Deflection AFM image. c – AFM-IR map with 1650 cm$^{-1}$ excitation. d– AFM-IR map with 1090 cm$^{-1}$ excitation. e – AFM-IR map with 1160 cm$^{-1}$ excitation. f – AFM-IR map with 1240 cm$^{-1}$ excitation. g – AFM-IR map with 1740 cm$^{-1}$ excitation. h – AFM-IR map with 2920 cm$^{-1}$ excitation. i – AFM-IR map with 3010 cm$^{-1}$ excitation.

Figure 3 shows AFM images of the sample and AFM-IR images for the same area. The selected wavelength values correspond to some of the main absorption bands. The map at 1650 cm$^{-1}$ matches the absorption peak of the Amide I band. In fixed cells, protein absorption is the main contribution to this spectral region, with some additional contribution from the tail of nucleic acid absorption around 1670 cm$^{-1}$. Therefore, the map is expected to be dominated by the distribution of proteins in the sample. The 1650 cm$^{-1}$ map appears to track closely the AFM height map and reproduces many of the filaments observed in the Height and Deflection maps. In addition, some particles or vesicles contribute strongly to the 1650 cm$^{-1}$ map, while others provide little or no contribution at this wavelength.

Several maps were run at the peak absorption frequency of the main bands from acyl lipids, 3010 cm$^{-1}$, 2920 cm$^{-1}$, 1740 cm$^{-1}$, 1240 cm$^{-1}$, 1160 cm$^{-1}$ and 1090 cm$^{-1}$ (Figure 3, panels, c, d, e, f, g, h). All the maps recorded at these frequencies are dominated by particles or vesicles and very little detail is associated to other subcellular structures. Only the map at 2920 cm$^{-1}$ shows some contributions from other parts of the cell. Most of these bands are co-localized within the same objects, suggesting that they arise mostly from the same molecules, confirming the assignment to acyl lipids. The size of most particles and vesicles observed in Figure 3 varies between 100 nm and 1500 nm. Most of them appear as uniform circular structures, either filled or with an inner cavity, suggesting that they result from the fixation of vesicles or micelles. A few of them appear as elongated objects, in some cases with an irregular inner structure, suggesting that they may result from the fixation of organelles, such as mitochondria or vacuoles. Comparison with the map at 1650 cm$^{-1}$ shows that spectral contributions from proteins are colocalized with those from lipids in some, but not all, of the particles.

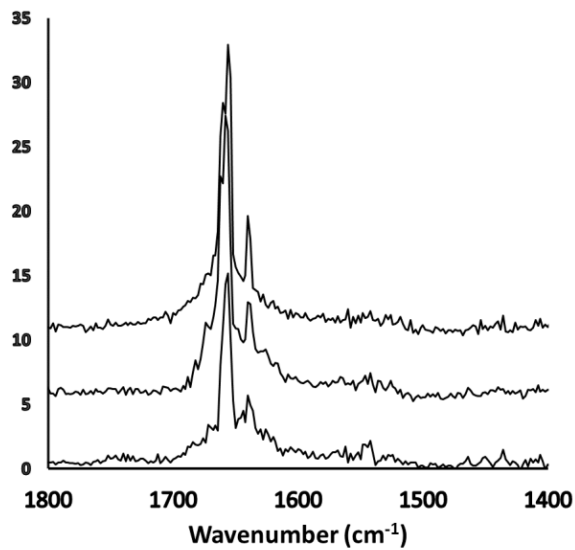

*Figure 4. Sharp peaks, at 1656 cm$^{-1}$ and 1638 cm$^{-1}$, overlying the Amide I band region. Spectra obtained from the measurement of cytoplasmic locations of Figure 2A.*

Inspection of individual spectra, such as the ones shown in Figure 2 (Positions 2 and 7), indicates the presence of sharp peaks that overlap the standard Lorentzian-Gaussian bands observed in FTIR spectra. These peaks are about 10 cm$^{-1}$ in FWHM and have a peak position that matches the one expected for the underlying standard FTIR feature. For example, Amide region spectra of the cytoplasmic portion of the cells, for which the Amide I region is shown in Figure 4, show the presence of two main peaks at 1638 cm$^{-1}$, 1656 cm$^{-1}$ and in some cases a shoulder at 1660 cm$^{-1}$. The bands match exactly the ones expected for the secondary structural elements of actin, which is composed predominantly of alfa helices (about 60%, absorbing at 1656 cm$^{-1}$) and of beta sheet (about 20%, absorbing at 1638 cm$^{-1}$) Similar sharp features are seen in the spectra of acyl bands at 2920 cm$^{-1}$ (not shown) and in the spectra of other materials. They are normally neglected and considered to be an artefact from exposure to high laser power. The origin of these bands will be one subject of the following discussion.

## Discussion

AFM-IR spectra and contact mode AFM-IR maps were recorded at different subcellular locations of fixed fibroblast cells. AFM-IR maps recorded at different wavelengths allow us to identify several subcellular structures, including micelles, lipid drops, vesicles and filaments. Resolution in the XY plane for AFM-IR experiments is approximately determined by the size of the tip, which is 30 nm in our experiments. In our experiments, we did not explore the resolution limits achievable in these samples. Nonetheless, the smallest structures observed in these measurements are about 50 nm in size, suggesting that, despite the complexity of a cellular sample, the achieved resolution is comparable to the one reported for systems with simpler structure and composition.

In most cases the resulting AFM-IR spectra display components similar to the ones observed in cellular spectra obtained with far-field focusing optics and FTIR technology. Specifically, we can observe several bands that are typically assigned to long-chain acyl lipids, including stretching C-H bond vibrations in the spectral range between 2800 cm$^{-1}$ and 3100 cm$^{-1}$, and headgroup vibrations between 1000 cm$^{-1}$ and 1800 cm$^{-1}$. Bands typically assigned to the backbone of proteins are also observed in the 1400 cm$^{-1}$ – 1700 cm$^{-1}$ range and in the 3200 cm$^{-1}$ – 3500 cm$^{-1}$ range. The assignments are supported by the observation that bands arising from the same species, such as the 2920 cm$^{-1}$ and 1740 cm$^{-1}$ bands from acyl lipids, must display the same spatial distribution in AFM-IR maps.

The assignment of bands to other macromolecules, such as nucleic acids and polysaccharides is uncertain. Absorption is observed in the spectral regions that are characteristic of polysaccharides, around 1000-1200 cm$^{-1}$. However, the relatively low spectral resolution, limited to 8 - 4 cm$^{-1}$ by the pulse length of the laser, and the signal-to-noise ratio afforded by the measurement do not allow resolving them from overlapping bands of proteins and lipids. Therefore, although a spectral contribution from molecules other than acyl lipids and proteins is expected, their bands cannot be univocally assigned in the present measurements.

When absorption at 1650 cm$^{-1}$, corresponding to the Amide I region, is mapped throughout the cell (Figure 3c), it appears to be associated to several particles of spherical shape, either micelles or vesicles that are also observed in AFM maps. In addition, the 1650 cm$^{-1}$ map reveals the presence of fibrillar structures distributed throughout the cytoplasm, about 100 nm to 1000 nm in diameter. Their absorption at 1650 cm$^{-1}$, characteristic of proteins, and their morphology suggest that the fibrils are components of the cytoskeleton, formed by the polymerization of G-actin monomers into F-actin. Actin is the most abundant protein in most eukaryotic cells [29] and it is expected to provide a dominant contribution to spectra from the cytoplasm. The fluorescence imaging experiments in Figure 1 confirm that the fixation preserves cellular actin filaments.

To our knowledge this is the first time that cytoskeletal structures are reported in AFM-IR maps of cells. Other investigators have measured AFM-IR maps of fixed eukaryotic cells with an excitation frequency of 1660 cm$^{-1}$, within the Amide I envelope. However, these authors did not report the fibril structure nor the particles that we observe in Figure 3a. [26]

It is notable that some spectra display a weak Amide II band relative to the corresponding Amide I. In FTIR transmission spectra of isotropic protein samples, including single cell spectra, the Amide I/Amide II intensity ratio is typically ~2. In the measurement reported in Figure 2, the ratio is always higher than 2. In some cases, the Amide II is barely observable (Positions 2 and 7). The lower values for the Amide I/Amide II ratio, closer to what is observed in single cell spectra, are observed in the location of lipid-rich particles and are probably due to the protein complement of these structures. The highest Amide I/Amide II ratios are obtained from measurements in locations where the lipid contribution is absent, either on particles that seem to be dominated by a protein contribution (Position 7) or in the cytoplasm. In several cytoplasmic locations, such as in Position 7, no other absorption bands are seen in the spectra except for a weak band at 1650 cm$^{-1}$.

Protein spectra with a weak or absent contribution from the Amide II band have been observed in AFM-IR measurements of purified fibrillar protein aggregates formed by lysozyme and the Josephin domain of Ataxin3. [30] In our sample, the presence of F-actin fibers raises the possibility that the unusual Amide I/Amide II ratio may be associated to these fibrils. However, the spectral region 1650 cm$^{-1}$ – 1600 cm$^{-1}$ is also affected by absorption from imine (C=N) bonds that are formed by the fixation process. It is therefore possible that the images at 1650 cm$^{-1}$ show a major contribution from the functional groups created during cross linking. The overlapping absorption from imines may be responsible for the unusually high Amide I/Amide II ratio observed in several cell locations. The contributions from amide and imine groups cannot be separated in the present experiment and additional experiments will be designed to address this point.

AFM-IR images of cells also reveal the presence of numerous particles, mostly in the size range between 100 nm and 1500 nm. Most of them can be associated to structures that are detectable in AFM Height and Deflection maps. Many of these particles have spherical appearance and most of them (with the exception of the one in Position 7) show a strong absorption from acyl lipid components. These spectral properties suggest that they originate from the fixation of vesicles or micelles, either already present in the live cell, such as endosomes and lipid droplets, or formed by the disruption of cellular components during fixation. Fluorescence imaging of fixed cells stained with BODIPY, which accumulates in lipid droplets, confirms that these structures survive the fixation. Presumably they correspond to the particles that in AFM-IR spectra display stronger absorption from acyl lipids with weaker contribution from protein bands (e.g. the particle in Figure 2, Position 5). Observation a sharp band at 3010 cm$^{-1}$ indicates that at least part of the lipid acyl chains possesses unsaturated C=C bonds. The intensity of this band relative to that of the bands at 2920 cm$^{-1}$ and 2985 cm$^{-1}$ can be used to estimate the degree of unsaturation. Comparison of the spectra for positions 1 and 4-6, shows that this is variable throughout the sample. The possibility to image lipid droplets in eukaryotic cells without any staining and the capability to assess the degree of unsaturation have also recently been demonstrated by Raman spectroscopy. [31, 32] Our work shows the possibility to obtain complementary spectroscopic information on these systems via AFM-IR measurements.

It is notable that only weak absorption from acyl lipid bands is observed throughout the cytoplasmic region, away from lipid droplets or vesicles. This is in contrast to what is observed in FTIR spectra of cells obtained with far field optics, both with fixed and with live samples. Although the latter measurements have relatively low, diffraction limited, resolution, they do allow in many cases collecting spectra of the cytoplasm separately from those of the nucleus. The cytoplasmic region typically shows strong acyl lipid absorption, [28] which has been attributed to the collective contribution from the Golgi and endoplasmic reticulum, and to various organelles such as mitochondria. In contrast, our AFM-IR measurements, both spectra and maps, show that at subcellular level acyl lipid absorption is mostly associated to lipid droplets and vesicles. It is not known if this is a consequence of membrane degradation during the fixation process or if it is due to reduced sensitivity of the AFM-IR technique to these structures.

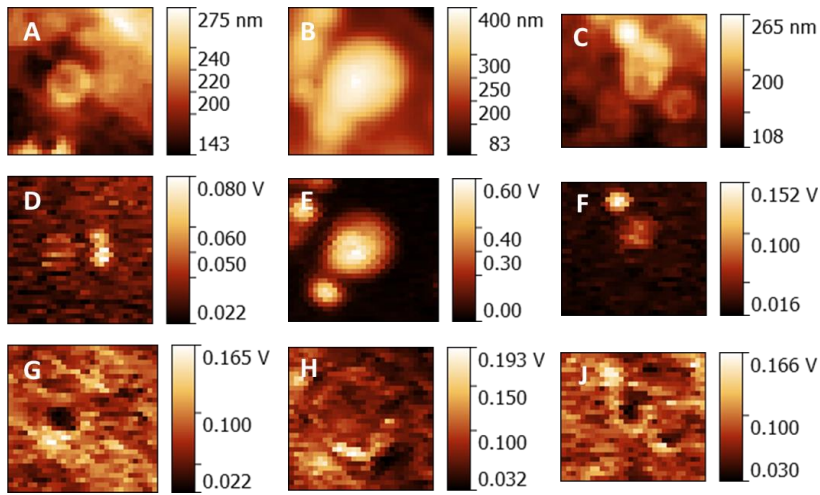

*Figure 5*. Details of three subcellular structures from Figure 3. A, B, C, Height maps. C, D, E, 2920 cm$^{-1}$ maps. 1650 cm$^{-1}$ maps.

In addition, particles with vesicular structure, displaying a shell enclosing a cavity, can be observed both in AFM Height maps and in AFM-IR maps (Figure 5). They are more obvious in maps recorded at 1650 cm$^{-1}$, indicating the presence of a large protein component. Some examples are shown in Figure 5, as details extracted from Figure 3. The vesicles may be endosomes or vesicles formed by the fragmentation of subcellular membranes or organelles. Their spectral features indicate that they retain much of the original complement of membrane proteins. Failure to observe extended structures that resemble the endoplasmic reticulum or the Golgi apparatus supports the proposition. Treatment of cells with paraformaldehyde is known to induce blebbing and fragmentation of the cell membrane and is used in the preparation of cell membrane vesicles. The process is used to produce vesicles with the composition of the membrane of origin but free from supporting cytoskeletal elements. We suggest that the paraformaldehyde treatment may be at the origin of the formation of some of these particles while the subsequent crosslinking reaction may contribute to their stabilization.

Some of the vesicle-like particles have a non-spherical shape and a non-uniform internal distribution (e.g. Figure 5 A, D, G and C, F, I), suggesting that they may be organelles or structures derived from their degradation. Several of them are about 1 µm in size. Mitochondria are 0.7 -1.5 µm in size and often display an elongated structure, making them a possible candidate. An alternative interpretation is that the particles are residual structures from the degradation of the endoplasmic reticulum or the Golgi apparatus. we show that the spatial resolution and spectral quality obtained in the measurements open the way to their in-depth characterization. Their exact assignment will be the subject of future investigations.

We observe sharp bands, about 10 cm$^{-1}$ in FWHM, in several locations of the sample. They are particularly obvious when overlying weak Amide bands, such as in Figure 4. These features are often reported in the AFM-IR spectra of various samples, such as polymers, but they are normally ignored or considered to be an unwanted effect of the use of high laser power. We propose that these bands are due to the non-linear response of the system when excited by moderately high laser power, as is the case in our experiments. Work in the non-linearity regime is used in photothermal imaging measurements the rely on the thermal lens effect. [33, 34] We propose that the same process is involved in signal generation in AFM-IR spectra when laser power is sufficiently high. The nonlinear regime is entered when the combined effect of laser power and tip pressure induce a transition or reaction in the sample, akin to what is observed in nanografting. [35] In the case of protein samples, the transition corresponds to the breakage of hydrogen bonds at peptide groups following excitation at the peak absorption of specific secondary structure elements. Other possible transitions that are possibly inducible under similar conditions are the melting of a polymer, a gel to liquid transition in phospholipids, or the unwinding of the DNA double helix.

The sharpness of the bands is due to the non-linear dependence from the electric field amplitude of the laser, which ensures that a phase transition or reaction occurs only in the proximity of the absorption peak. [33] Therefore, a consequence of non-linearity is that the achievable resolution is higher than the one allowed by the duration of the laser pulse in a conventional AFM-IR experiment. The result, in the case of a polypeptide, as in Figure 4, is a spectrum in which the individual components of the Amide I band are fully resolved. An improvement in the apparent spectral resolution of Amide I band components in AFM-IR spectra of proteins has already been reported [36] but an explanation has not yet been found.

Using a detection scheme based on the deflection of an AFM tip to study a process in a nonlinear photothermal regime allows us to achieve similar sensitivity as photothermal imaging based on the thermal lens effect. However, in contrast to the latter scheme, the response is only a function of the imaginary part of the refractive index and allows us to perform accurate spectroscopic measurements. In addition, an AFM-IR setup allows us to achieve a spatial resolution comparable to the size of the AFM tip. This is at least one order of magnitude better than what is allowed by a laser deflection scheme, for which a spatial resolution just below 1 µm has been reported. [37] We can state that, by operating in the nonlinear regime, an AFM-IR experiment can take advantage of both worlds, and allow high sensitivity measurements with high spatial resolution.

## Conclusions

We have tested the capability of AFM-IR to perform subcellular studies of fixed fibroblast cells. We show that the technique allows us to image subcellular structures that include lipid droplets, vesicles, organelles and cytoskeletal components with intrinsic contrast that relies on their IR absorption. The spectroscopic properties of these structures are accessible via single point measurements. This work opens the way to studies that will take advantage of AFM-IR to investigate the properties of individual organelles and other subcellular entities.

In addition, we demonstrate the possibility to perform AFM-IR measurements by using moderately high power and operating in the non-linearity regime region of the photothermal response. This is major advance in our control of the AFM-IR technique, allowing high-sensitivity and high resolution studies of the sample. One major advantage is that it does not require the use of a flat gold optical substrate for plasmonic enhancement, as in resonance AFM experiments, providing greater flexibility in optical design. [38] Furthermore, non-linear effects in photothermal spectroscopy rely on the onset of a phase transition or reaction, and AFM-IR can be used to detect and study such transition. The implication is that an AFM-IR measurement can in principle be used as the detection scheme of a calorimetric experiment for nanometric samples.

The challenge for the near future will be the extension of these measurement capabilities to living cells. This milestone, when achieved, will lead to a whole new approach in the investigation of cellular biochemistry.

**Acknowledgments**

The research was performed using equipment purchased in the frame of the project co-funded by the Małopolska Regional Operational Programme Measure 5.1 Krakow Metropolitan Area as an important hub of the European Research Area for 2007-2013, project No. MRPO.05.01.00-12-013/15.

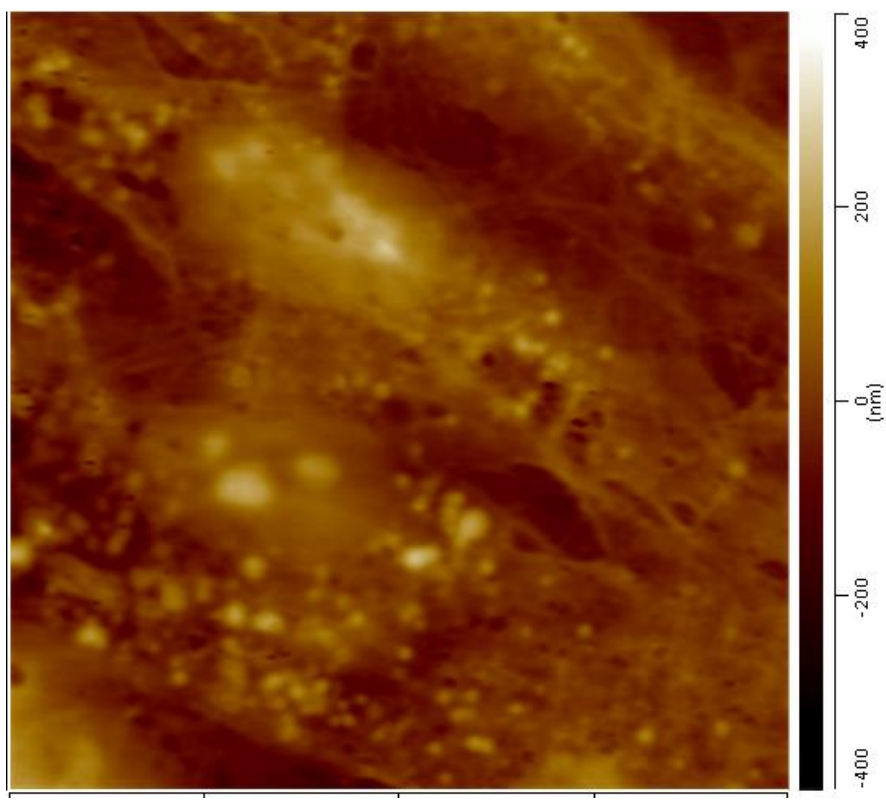

**Supplementary Figure 1**. Overview of sample area in an AFM Height map